\documentclass[prb,twocolumn,superscriptaddress,showpacs,amsmath,amssymb,longbibliography]{revtex4-2}
\usepackage{amsmath, amsfonts, amssymb, bm}
\usepackage{float}
\usepackage{mathtools}
\usepackage[inline]{enumitem}
\usepackage{graphicx}
\usepackage{xcolor}
\usepackage[breaklinks,hypertexnames=false]{hyperref}
\hypersetup{colorlinks,linkcolor={magenta},citecolor={blue},urlcolor={blue}}
\usepackage[capitalize]{cleveref}

\begin{document}
	
\title{Spin-polarized Josephson current induced by inhomogeneous altermagnetic interlayers}
	
\newcommand{\tianjin}{Center for Joint Quantum Studies, Tianjin Key Laboratory of Low Dimensional Materials Physics and Preparing Technology, Department of
Physics, Tianjin University, Tianjin 300354, China}
	
\newcommand{\nagoya}{Department of Applied Physics, Nagoya University, 464-8603 Nagoya, Japan}
	
\newcommand{\Okayama}{Faculty of Environmental Life, Natural Science and Technology, Okayama University, 700-8530 Okayama, Japan}
	
\newcommand{\Uppsala}{Department of Physics and Astronomy, Uppsala University, Box 516, S-751 20 Uppsala, Sweden}
	
\newcommand{\uam}{Department of Theoretical Condensed Matter Physics, Universidad Aut\'onoma de Madrid, 28049 Madrid, Spain}
\newcommand{\ifimac}{Condensed Matter Physics Center (IFIMAC), Universidad Aut\'onoma de Madrid, 28049 Madrid, Spain}
\newcommand{\inc}{Instituto Nicol\'as Cabrera, Universidad Aut\'onoma de Madrid, 28049 Madrid, Spain}

\author{Wenjun Zhao}
\affiliation{\tianjin}

\author{Yuri Fukaya}
\affiliation{\Okayama}
	
\author{Pablo Burset}
\affiliation{\uam}
\affiliation{\ifimac}
\affiliation{\inc}
	
\author{Jorge Cayao}
\affiliation{\Uppsala}

\author{Yukio Tanaka}
\affiliation{\nagoya}

\author{Bo Lu}
\affiliation{\tianjin}

\date{\today}

\begin{abstract}
{The pursuit of dissipationless spin supercurrents is a central theme in superconducting spintronics. We propose a field-free Josephson junction using an inhomogeneous altermagnetic interlayer with in-plane N\'{e}el vectors. We show that the current-phase relation and the critical Josephson current are highly sensitive to the misorientation angle between the altermagnetic layers' N\'{e}el vectors. Specifically, at a $\pi$ misorientation with equal layer thicknesses the spatial oscillations of the superconducting pair amplitude, governed by the center-of-mass momentum, undergo mutual cancellation. This compensation suppresses individual layer pair-breaking, significantly enhancing the critical current and eliminating $0$-$\pi$ transitions. Furthermore, the non-collinear alignment of the N\'{e}el vectors facilitates the emergence of a net spin-polarized Josephson current. This spin current serves as a distinct signature of spin-triplet pair correlations, generated by the spin-dependent momentum shifts inherent to the altermagnetic exchange field. Our results establish a highly tunable, field-free platform for the realization of dissipationless spintronic devices.}
\end{abstract}
\maketitle

\section{Introduction}
The Josephson effect in ferromagnet (FM)-superconductor (SC) hybrids has long been a subject of research for exploring novel quantum transport phenomena~\cite{Golubov04,Buzdin05,Bergeret05} and has enabled the entire field of superconducting spintronics~\cite{LinderR2015,Eschrig_2015,mel2022superconducting}. In particular, Josephson junctions consisting of inhomogeneous FMs have been extensively studied in the last two decades~\cite
{Bergeret2001,Bergeret2001prb,Krivoruchko2001prb,Koshina2001,Golubov2002,Barash2002,Blanter04,Buzdin2006prb,Crouzy2007,asano2007prl,asano2007prb,Sperstad2008,Alidoust10,Trifunovic11,Shomali_2011,Linder14,Halterman15,Lu2020Jan}. The Josephson current is known to exhibit $0$-$\pi$ transitions when the orientations of the FMs are parallel~\cite{Aarts2001Mar,Golubov04}, but with critical currents that are much smaller than for an antiparallel arrangement of FMs~\cite{Bergeret2001,Bergeret2001prb,Krivoruchko2001prb}. Furthermore, for non-collinear FMs, long-ranged spin-triplet supercurrents can be generated, resulting in the experimentally observed long-range proximity effect~\cite{longrangeExp,Birge2010Mar,Buzdin05,Bergeret05,Volkov2003prl,BVE2003prb,Houzet2007}. While all these advances are indeed promising, 
the finite net magnetization and unavoidable stray fields of FMs introduce serious challenges for maintaining the integrity of the superconducting condensate in the junction region~\cite{blamire2014interface,Paschoa2020}.

\begin{figure}[t]
\centering
\includegraphics[width=8.7cm]{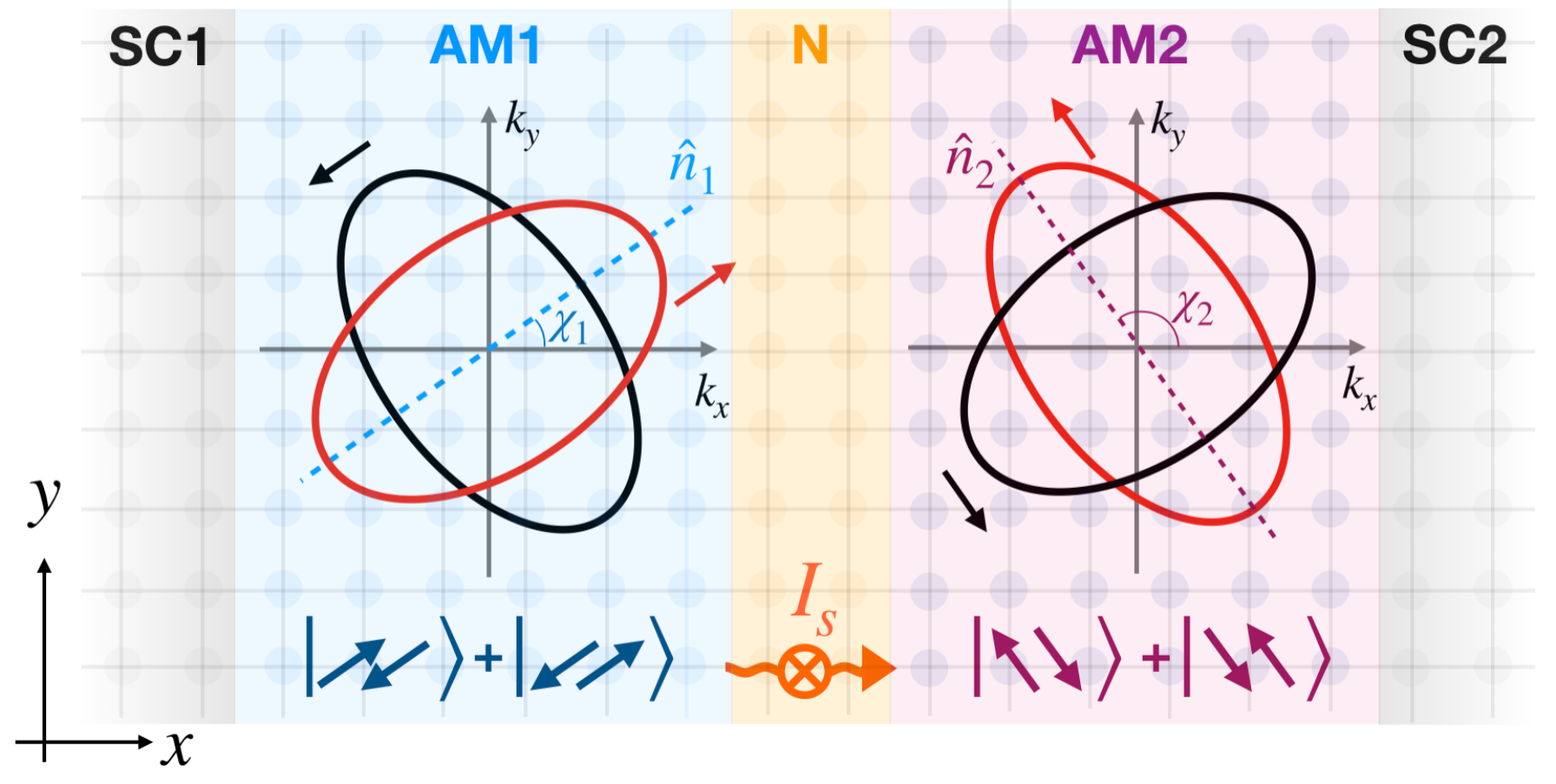}
\caption{Sketch of a two-dimensional SC1-AM1-N-AM2-SC2 Josephson junction on the tight-binding lattice. The black and red ellipses represent the two spin species of the altermagnetic spin bands in momentum space and $\hat{n}_{1(2)}=(\cos\chi_{1(2)},\sin\chi_{1(2)},0)$ denotes the in-plane N\'{e}el vector in AM1 (AM2), i.e., the tilt angle of the spin-band lobe relative to the $x$ axis. The misorientation of spin-triplet pair amplitude with non-collinear AMs drives an out-of-plane spin supercurrent $I_s$.}
\label{fig1}
\end{figure}

The effects of FMs can be mitigated by using instead the recently discovered altermagnets (AM) ~\cite{landscape22,LiborPRX22,MazinPRX22,Bailing,Krempasky2024,Songreview2025,FukayaJPCM2025}.
In ferromagnetic Josephson junctions, macroscopic net magnetization creates stray magnetic fields and orbital pair-breaking that severely suppress the superconducting condensate. 
By contrast, momentum-dependent nonrelativistic spin splitting had previously been identified in various collinear antiferromagnets~\cite{spin-splitting1,spin-splitting2,spin-splitting3,spin-splitting4,spin-splitting5,spin-splitting6,spin-splitting7}.
More recently, materials with zero net magnetization but finite momentum-dependent spin splitting have been unified and formally classified as AMs based on spin-group symmetry principles~\cite{liu1}.
AMs thus enable a completely field-free platform that protects superconductivity while generating intriguing phenomena, representing a fertile ground for exploring emergent superconducting effects which has already attracted great interest~\cite{Brekke23,Zhongbo23,CCLiu1,mondal2024,Schaffer2024,Giil_2024,Canoprl24,Banerjeediode24,Anjishnu2024,Wangjian2024,soori2024,nagae2024,Sukhachov2024,Chourasia2025,FukayaJPCM2025,Song2025prl,AmartyaL25,GiBaik2025,Chakraborty25con,annica2025,Chakraborty25,parshukov2025,Xiaji2025,Cayaoprl26,Bolu2026,fukayaprl2026,Heinsdorf2026,Ohidul2026}. Of particular importance is the potential of AMs to drive novel phenomena in Josephson junctions, including $0$-$\pi$ transitions~\cite{Ouassou23,Beenakker23,zhang2024,heras2026,Darvishi26,mondal2026NJE}, phase-shifted Andreev levels~\cite{Beenakker23}, exotic $\varphi $-junctions~\cite{Bo2024,sun2024,fukaya2024}, high-harmonic current-phase relations~\cite{Bo2024,sun2024,fukaya2024,Wenjun25,Cheng24}, and Josephson diode effects~\cite{chengdiode24,Lovy2025,Boruah2025,mondal2026NJE}. Notably, the generation and manipulation of superconducting spin currents in these hybrid structures has emerged as a pivotal theme, offering a unique platform for dissipationless spintronics~\cite{Lun-Hui26,lichuang2026,Monkman2026PRX,fu2026spincurrent}. However, current studies mainly focus on homogeneous interlayers, leaving the physics of composite altermagnetic structures and their potential for tailoring superconducting spin currents largely unexplored. 
In a single-layer junction, the uniform orientation of the AM exchange field precludes spin torques, resulting in a strictly vanishing spin supercurrent.
By contrast, multilayer Josephson junctions, particularly double-FM structures, are known to host rich quantum interference phenomena.
Extending the multilayer paradigm to AMs is therefore a natural question, specially since real AM materials can exhibit spatial inhomogeneities such as twin boundaries.
These considerations motivate the study of double-AM junctions and raise the question of whether their anisotropic spin splitting can give rise to anomalous Josephson effects.

In this paper, we investigate a two-dimensional hybrid system consisting of two AM layers sandwiched between conventional superconducting leads, as illustrated in Fig. \ref{fig1}. 
Each layer possesses a uniform altermagnetic strength and an in-plane N\'{e}el vector, with orientations $\chi_1$ and $\chi_2$ that can be tuned independently. While similar junction structures have been studied in the context of tunneling magnetoresistance~\cite{QingFengtmr}, these investigations have focused on non-superconducting configurations.
Using the lattice Green’s function method, we calculate the Josephson effect in this system. We find that at a $\pi$ misorientation between $\chi_1$ and $\chi_2$, the oscillations of the superconducting pair amplitude driven by the center-of-mass momenta perfectly cancel in layers of equal thickness. This compensation suppresses pair-breaking, significantly enhances the Josephson current, and eliminates $0$-$\pi$ transitions. Furthermore, we demonstrate that noncollinear magnetic configurations generate a net, tunable spin-polarized Josephson current, indicating that the altermagnetic field facilitates the conversion of spin-singlet pairs into spin-triplet ones. Importantly, despite the sign-changing $d$-wave symmetry of AM spin splitting in momentum space, our work unambiguously shows that a robust net spin supercurrent persists after momentum-angle integration across the Fermi surfaces. Finally, the generated spin supercurrent exhibits a high degree of tunability and nonreciprocity. Its magnitude, sign, and current-phase relationship can be continuously manipulated by tuning the individual layer orientation or the junction length, enabling perfect nonreciprocal spin-polarized transport. Our results thus introduce a highly tunable, field-free platform for the realization of dissipationless spin currents.

The paper is organized as follows: In Sec.~\ref{sec2}, we introduce our model and formalism. Sec.~\ref{sec3} contains the numerical results for the orientation dependence of Josephson charge current. We study in Sec.~\ref{sec4} the emergence of the Josephson spin current when the N\'{e}el vectors are non-collinear. Our conclusions are given in Sec.~\ref{sec5}.

\section{Model and Formalism}
\label{sec2}

We consider a two-dimensional Josephson junction where two SCs are connected by two AMs separated by a normal (N) region; the SC1-AM1-N-AM2-SC2 junction depicted in Fig.~\ref{fig1}.
Translational invariance along the $y$ direction, with $%
k_{y}$ the conserved wave vector, allows us to decompose the two-dimensional system into multiple $k_{y}$-resolved one-dimensional lattice chains.
The tight-binding Hamiltonian of the junction is given by~\cite{fukaya2024}
\begin{align}
\check{H}_{\text{TB}} =&\sum_{i,k_{y}}\left[ \hat{\psi}_{i,k_{y}}^{\dag }%
\check{u}_{k_{y}}\hat{\psi}_{i,k_{y}}+\hat{\psi}_{i,k_{y}}^{\dag }\check{T}%
_{01,k_{y}}\hat{\psi}_{i+1,k_{y}}\right.   \notag \\
&\left. + \, \hat{\psi}_{i+1,k_{y}}^{\dag }\check{T}_{10,k_{y}}\hat{\psi}%
_{i,k_{y}}\right] ,
\end{align}%
where $i$ labels the lattice sites along the $x$ direction and $\hat{\psi}%
_{i,k_{y}}=(c_{i,k_{y},\uparrow },c_{i,k_{y},\downarrow
},c_{i,-k_{y},\uparrow }^{\dag },c_{i,-k_{y},\downarrow }^{\dag })^{T}$ is
the Nambu field operator at site $i$. 
The on-site potential, left-hopping, and right-hopping matrices are respectively denoted by $\check{u}_{k_{y}}$, $\check{T}_{01,k_{y}}$, and $\check{T}_{10,k_{y}}=(\check{T}_{01,k_{y}})^{\dagger }$. Below, the Pauli matrices $\hat{s}_{j}$ and $\hat{\tau}_{j}$ ($j=0,x,y,z$) act on spin and Nambu space, respectively.

In the SC leads, the hopping terms are uniform with $\check{T}_{01,k_{y}}^{%
\text{SC}}=\check{T}_{10,k_{y}}^{\text{SC}}=-t_{1}\hat{\tau}_{z}\equiv
\check{T}$, where $t_{1}$ is the nearest-neighbor hopping energy along the $x
$ direction. The on-site term in the SCs reads
\begin{equation}
\check{u}_{k_{y}}^{\text{SC}}=\varepsilon _{k_{y}}\hat{\tau}_{z}-\Delta \hat{%
s}_{y}\hat{\tau}_{y},
\end{equation}%
where $\varepsilon _{k_{y}}=-\mu +4t_{1}-2t_{1}\cos k_{y}$ represents the
normal-state dispersion relative to the chemical potential $\mu $. The
superconducting pair potential is $\Delta =\Delta _{0}e^{i\varphi }$ in SC1
and $\Delta =\Delta _{0}$ in SC2, with $\varphi $ being the macroscopic
Josephson phase difference. The gap temperature dependence is given by $%
\Delta (T)=\Delta _{0}\tanh (1.74\sqrt{T_{c}/T-1})$, with $\Delta
_{0}=3.53k_{B}T_{c}/2$, and $T$, $k_{B}$, and $T_{c}$ respectively being the temperature,
Boltzmann constant, and critical temperature. 
For the N region, we
fix the number of lattice sites to $N=2$ without loss of generality. The
corresponding tight-binding parameters are $\check{u}_{k_{y}}^{\text{N}%
}=\varepsilon _{k_{y}}\hat{\tau}_{z}$ and $\check{T}_{01,k_{y}}^{\text{N}}=%
\check{T}_{10,k_{y}}^{\text{N}}=\check{T}$. In the AM layers (AM1 and AM2,
indexed by $\alpha \in \{1,2\}$), the number of lattice sites are $N_{1}$
for AM1 and $N_{2}$ for AM2. 
The on-site and hopping matrices in AM$\alpha $
are given by
\begin{align}
\check{u}_{k_{y}}^{\text{AM}\alpha }={}& \varepsilon _{k_{y}}\hat{\tau}%
_{z}+J_{2}^{\alpha }\mathcal{\check{S}}(\chi _{\alpha })\cos k_{y}, \\
\check{T}_{01,k_{y}}^{\text{AM}\alpha }={}& \left( \check{T}_{10,k_{y}}^{%
\text{AM}\alpha }\right) ^{\dagger }=\left[ \frac{J_{1}^{\alpha }}{2i}\sin
k_{y}-\frac{J_{2}^{\alpha }}{2}\right] \mathcal{\check{S}}(\chi _{\alpha })+\check{T},
\end{align} 
where $J_{1}^{\alpha }=2J\sin (2\chi _{\alpha })$ and $J_{2}^{\alpha
}=2J\cos (2\chi _{\alpha })$ respectively represent the $d_{xy}$-wave and $d_{x^{2}-y^{2}}
$-wave components of the altermagnetic exchange interaction of strength $J$. Here, we consider that
both AM layers have the same AM
strength, and $\chi _{\alpha }$ is the orientation angle of the AM lobe relative to the $x$ axis.
For $\chi_{\alpha} = n\pi/2$ ($n=0,1,2,\dots$), the exchange splitting exhibits pure $d_{x^2-y^2}$-wave symmetry, whereas for $\chi_{\alpha} = (2n+1)\pi/4$, it exhibits pure $d_{xy}$-wave symmetry. Assuming the N\'{e}el vector lies along the $\chi_\alpha$ direction within the $x$-$y$ plane,
i.e., it co-rotates with the crystal axes as when crystalline anisotropy pins the easy axis,
the AM exchange matrix in Nambu space is explicitly defined as
\begin{eqnarray}
\mathcal{\check{S}}(\chi _{\alpha }) &=&%
\begin{pmatrix}
\hat{s}(\chi _{\alpha }) & 0 \\
0 & -\hat{s}(-\chi _{\alpha })%
\end{pmatrix},
\end{eqnarray}%
with $\hat{s}(\chi _{\alpha })=\cos \chi_\alpha \hat{s}_{x}+\sin \chi_\alpha \hat{s}_{y}$. 
%{\color{blue}Here $\chi_\alpha$ denotes the common rotation angle of the altermagnetic lobes and the N\'eel vector, i.e., the N\'eel vector is assumed to co-rotate with the crystal axes, as when crystalline anisotropy pins the easy axis.} 
Finally, the hopping matrix at the SC/AM interfaces is modeled as $\check{T}%
_{I}=t_{\text{int}}\check{T}$, where $t_{\text{int}}$ characterizes the
interface transparency. The hoppings across the AM/N interfaces are set to $%
\check{T}$.

Using the standard recursive Green's function method~\cite%
{FURUSAKI1994214,Umerski97,Asano2001,san2013multiple,Yada2017,FukayaPRB2020,fukaya2022npj,Yada2023,fukaya2024}%
, 
the Josephson charge current $I_c$ and $z$-polarized spin current $I_s$ are calculated inside the central N region as follows
\begin{equation}
I_{c\left( s\right) }=ieT\sum\limits_{\omega _{n},k_{y}}\text{Tr}[\hat{s}%
_{0\left( z\right) }\hat{\tau}_{z}\check{T}(\check{G}_{01,\omega _{n},k_{y}}-%
\check{G}_{10,\omega _{n},k_{y}})].
\label{current}
\end{equation}%
Here, $\omega _{n} = (2n + 1)\pi k_B T$ are
Matsubara frequencies, $\check{G}_{01,\omega _{n},k_{y}}$ and $\check{G}_{10,\omega
_{n},k_{y}}$ are the nonlocal part of the Matsubara Green's function between two
adjacent layers inside the N regions, given by $\check{G}_{01,\omega
_{n},k_{y}}=\check{G}_{\omega _{n},k_{y}}^{\text{L}}\check{T}\check{G}%
_{11,\omega _{n},k_{y}}$, and $\check{G}_{10,\omega _{n},k_{y}}=\check{G}%
_{\omega _{n},k_{y}}^{\text{R}}\check{T}^{\dag }\check{G}_{00,\omega
_{n},k_{y}}$. $\check{G}_{00,\omega _{n},k_{y}}$ and $\check{G}_{11,\omega
_{n},k_{y}}$ are the local Green's function in the N region, namely, 
\begin{eqnarray}
\check{G}_{00,\omega _{n},k_{y}} &=&[(\check{G}_{\omega _{n},k_{y}}^{\text{L}%
})^{-1}-\check{T}\check{G}_{\omega _{n},k_{y}}^{\text{R}}\check{T}]^{-1}, \\
\check{G}_{11,\omega _{n},k_{y}} &=&[(\check{G}_{\omega _{n},k_{y}}^{\text{R}%
})^{-1}-\check{T}\check{G}_{\omega _{n},k_{y}}^{\text{L}}\check{T}]^{-1},
\label{greenpole}
\end{eqnarray}%
where $\check{G}_{\omega _{n},k_{y}}^{\text{L}}$ and $\check{G}_{\omega
_{n},k_{y}}^{\text{R}}$ are given by%
\begin{eqnarray}
\check{G}_{\omega _{n},k_{y}}^{\text{L}} &=&[i\omega _{n}-\check{u}_{k_{y}}^{%
\text{N}}-\check{T}\mathcal{\check{G}}_{\omega _{n},k_{y}}^{\text{(N}_{1}%
\text{)}}\check{T}]^{-1}, \\
\check{G}_{\omega _{n},k_{y}}^{\text{R}} &=&[i\omega _{n}-\check{u}_{k_{y}}^{%
\text{N}}-\check{T}\mathcal{\check{K}}_{\omega _{n},k_{y}}^{\text{(N}_{2}%
\text{)}}\check{T}]^{-1}.
\end{eqnarray}%
%{\color{blue}Here, $i\omega _{n} = i(2n + 1)\pi k_B T$ are Matsubara frequencies.} 
Finally, $\mathcal{\check{G}}_{\omega _{n},k_{y}}^{\text{(N}_{1}\text{)}}$ ($%
\mathcal{\check{K}}_{\omega _{n},k_{y}}^{\text{(N}_{2}\text{)}}$) is the
surface Green's function at the right (left) edge of AM1 (AM2), which is obtained recursively as
\begin{eqnarray}
\mathcal{\check{G}}_{\omega _{n},k_{y}}^{\left( j\right) } &=&\left[ i\omega
_{n}-\check{u}_{k_{y}}^{\text{AM},1}-\check{T}_{10,k_{y}}^{\text{AM},1}%
\mathcal{\check{G}}_{\omega _{n},k_{y}}^{\left( j-1\right) }\check{T}%
_{01,k_{y}}^{\text{AM},1}\right] ^{-1}, \\
\mathcal{\check{K}}_{\omega _{n},k_{y}}^{(j)} &=&\left[ i\omega _{n}-\check{u%
}_{k_{y}}^{\text{AM},2}-\check{T}_{01,k_{y}}^{\text{AM},2}\mathcal{\check{K}}%
_{\omega _{n},k_{y}}^{(j-1)}\check{T}_{10,k_{y}}^{\text{AM},2}\right]^{-1}.
\end{eqnarray}%
The left (right) superconducting lead is coupled as a surface
Green's function $\check{g}_{\omega _{n},k_{y}}$ ($\check{k}_{\omega
_{n},k_{y}}$) at the edge of SC1 (SC2) as follows
\begin{eqnarray}
\mathcal{\check{G}}_{\omega _{n},k_{y}}^{(1)} &=&[i\omega _{n}-\check{u}%
_{k_{y}}^{\text{AM},1}-\check{T}\check{g}_{\omega _{n},k_{y}}\check{T}]^{-1},
\\
\mathcal{\check{K}}_{\omega _{n},k_{y}}^{(1)} &=&[i\omega _{n}-\check{u}%
_{k_{y}}^{\text{AM},2}-\check{T}\check{k}_{\omega _{n},k_{y}}\check{T}]^{-1}.
\end{eqnarray}%
The surface Green's functions $\check{g}_{\omega _{n},k_{y}}$ and $\check{k}%
_{\omega _{n},k_{y}}$ are calculated by the M\"{o}bius transformation
following Ref.~\cite{Umerski97}.

For the numerical simulations, we set $t_1=1$, $\mu =1.5t_1$, and an interface
transparency $t_{int}=t_1$. The critical temperature for both SCs is fixed at $k_{B}T_{c}=0.01t_1$, and we set the temperature $T=0.025T_{c}$ to represent
the low-temperature limit. Under these conditions, the superconducting coherence
length is $\xi \approx 100a$~\cite{fukaya2024}, with $a$ the lattice unit. We define the junction length as $%
L=(N_{1}+N+N_{2})a$. For simplicity, we assume equal AM layer thicknesses ($N_1 = N_2$), and maintain $N=2$ throughout our analysis. In our calculation, $L$ is limited to a maximum of $70a$ without disorder. The investigation of diffusive transport arising from impurity scattering in longer junctions is left for future work.

\begin{figure*}[htbp]
\centering
\includegraphics[width=14cm]{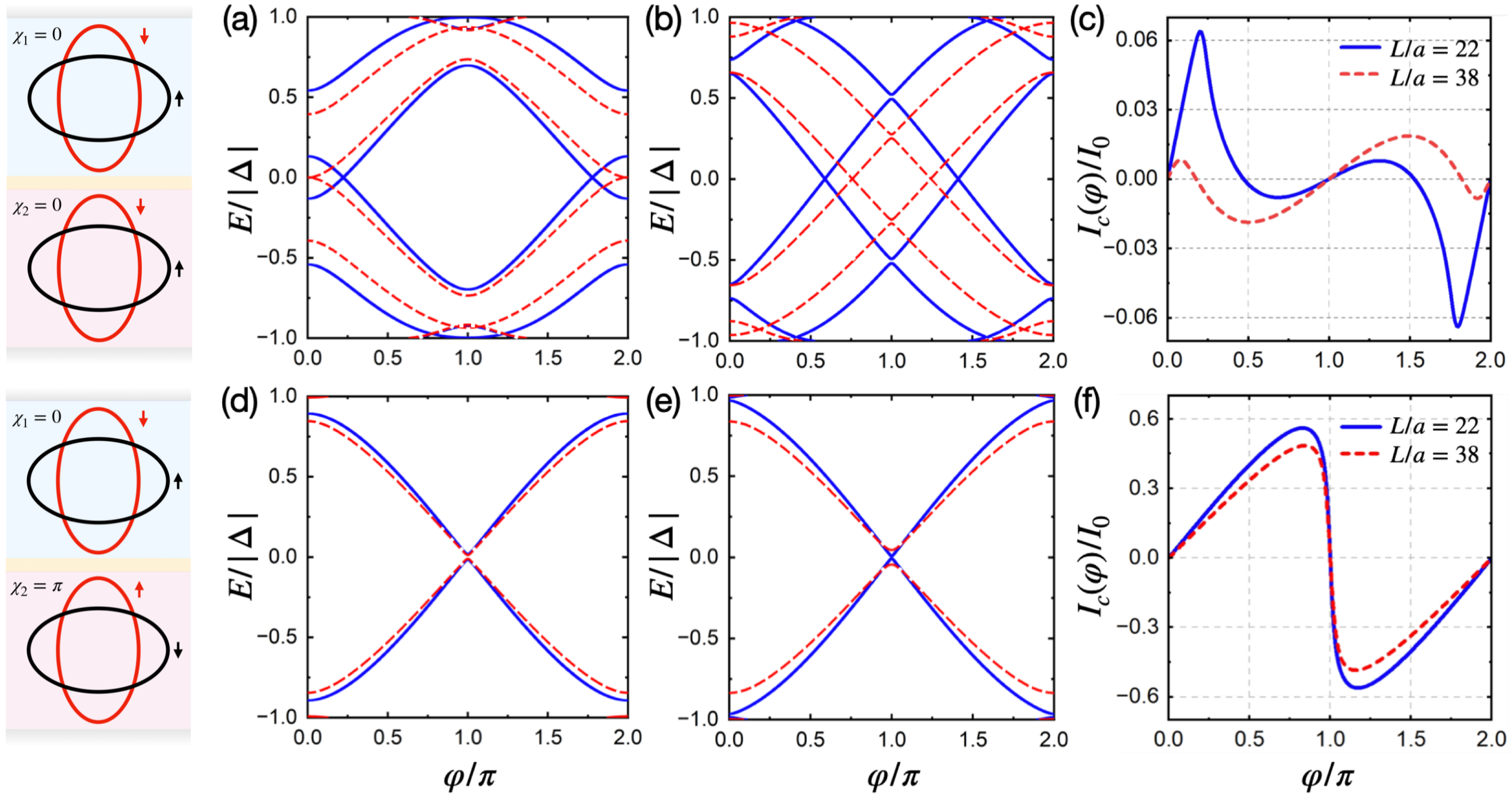}
\caption{(a,b) Andreev spectrum as a function of the superconducting phase $\varphi$ for  $k_y=0$ (a) and $k_y=0.1%
\protect\pi$ (b), both at $J=0.25t$. (c) Total charge Josephson current versus $\varphi$; here, $I_0$ is the maximum Josephson current of a two-dimensional SC-N-SC junction with a two-site normal region. Leftmost panel depicts the Fermi surfaces of the AMs in (a-c), with the Neel vectors having relative orientations of $\protect\chi_1=0$ and $\protect\chi_2=0$. Blue solid and red dashed curves denote junction lengths of $L=22a$ and $L=38a$, respectively. (d-f) Same as (a-d) but for $\protect\chi_1=0$ and $\protect\chi_2=\pi$, shown in the leftmost panel of the bottom row. 
} 
\label{fig2}
\end{figure*}

\section{Josephson charge current}
\label{sec3}
We first analyze the Andreev bound states (ABSs) and the current-phase relation (CPR) of the Josephson junction by comparing two specific configurations of the AM layers. The ABSs are obtained from the poles of Green's function in the middle normal region N using Eq.~(\ref{greenpole}). The CPR is calculated from Eq.~(\ref{current}). We fix the orientation of the left AM layer at $\chi_1 = 0$ while setting the right AM layer to $\chi_2 = 0$ [parallel configuration, see Figs.~\ref{fig2}(a)-(c)] and $\chi_2 = \pi$ [antiparallel configuration, see Figs.~\ref{fig2}(d)-(f)]. For the parallel configuration, we examine the ABSs for fixed $k_y = 0$ and $k_y = 0.1\pi$ as a function of the superconducting phase difference $\varphi$. The ABSs exhibit significant splittings and can develop crossings at zero energy with phase differences other than $\pi$ [Figs.~\ref{fig2}(a)(b)]. Such behavior can be understood from the fact that the pair amplitude induced in proximized AMs acquires a finite center-of-mass momentum~\cite{zhang2024,finite-momentum2,finite-momentum4,finite-momentum6,finite-momentum1,finite-momentum3,finite-momentum5}. The magnitude of the center-of-mass momentum changes for each $k_y$ channel due to the varying spin-splittings in the Brillouin zone~\footnote{The zero-energy crossings away from $\pi$ are similar to those seen in Josephson junctions with Rashba spin-orbit coupling under magnetic field~\cite{yokoyama2013josephson,PhysRevB.91.024514,PhysRevB.89.195407, Lu2015Aug, PhysRevB.96.205425,cayao2018andreev,PhysRevX.9.011010,baldo2023zero,PhysRevLett.123.117001,PhysRevB.109.L081405,PhysRevB.105.054504,79tj-c3y4}, where they were shown to induce sharp transitions in the current-phase curves.}. Consequently, the momentum-resolved components of the Josephson current exhibit $0$-$\pi$ transitions. In our analysis, the total current is normalized by $I_0$, which represents the maximum Josephson current of an equivalent SC-N-SC junction with a two-site normal interlayer. As shown in Fig.~\ref{fig2}(c), the total CPR can undergo a $0$-$\pi$ transition as the junction length $L$ changes, consistent with previous studies on SC-AM-SC junctions~\cite{Ouassou23,Beenakker23,zhang2024,Bo2024,sun2024,fukaya2024}. Furthermore, the CPR with exotic skewness [blue curve in Fig.~\ref{fig2}(c)] can emerge because the momentum-resolved ABSs exhibit competing $0$- or $\pi$-type characteristics across different transport channels. 

By contrast, for the antiparallel configuration ($\chi_R = \pi$) the splitting of the ABSs disappear for arbitrary $k_y$ [Figs.~\ref{fig2}(d)(e)]. These states instead contribute to a standard $0$-type CPR. As a result, the total CPR reverts to the conventional form observed in standard SC-N-SC junctions~\cite{Kulik1978SJLT}, as shown in Fig.~\ref{fig2}(f). Indeed, we find that whenever the relative orientation between the two layers is $\Delta\chi = \pi$, the system effectively compensates for the altermagnetic orderings. This phenomenon is analogous to the behavior of antiparallel ferromagnets in a Josephson junction, where the CPR is restored to its conventional form~\cite{Bergeret2001,Bergeret2001prb,Krivoruchko2001prb}. Because the N\'{e}el vectors of the two AM layers are antiparallel in this configuration, the spatial oscillations of the superconducting pair amplitude, governed by the center-of-mass momenta, undergo mutual cancellation. Consequently, the pair-breaking effects inherent to the individual layers are suppressed, and all $k_y$ channels display a $0$-type CPR.

\begin{figure}[tbp]
\centering
\includegraphics[width=8.8cm]{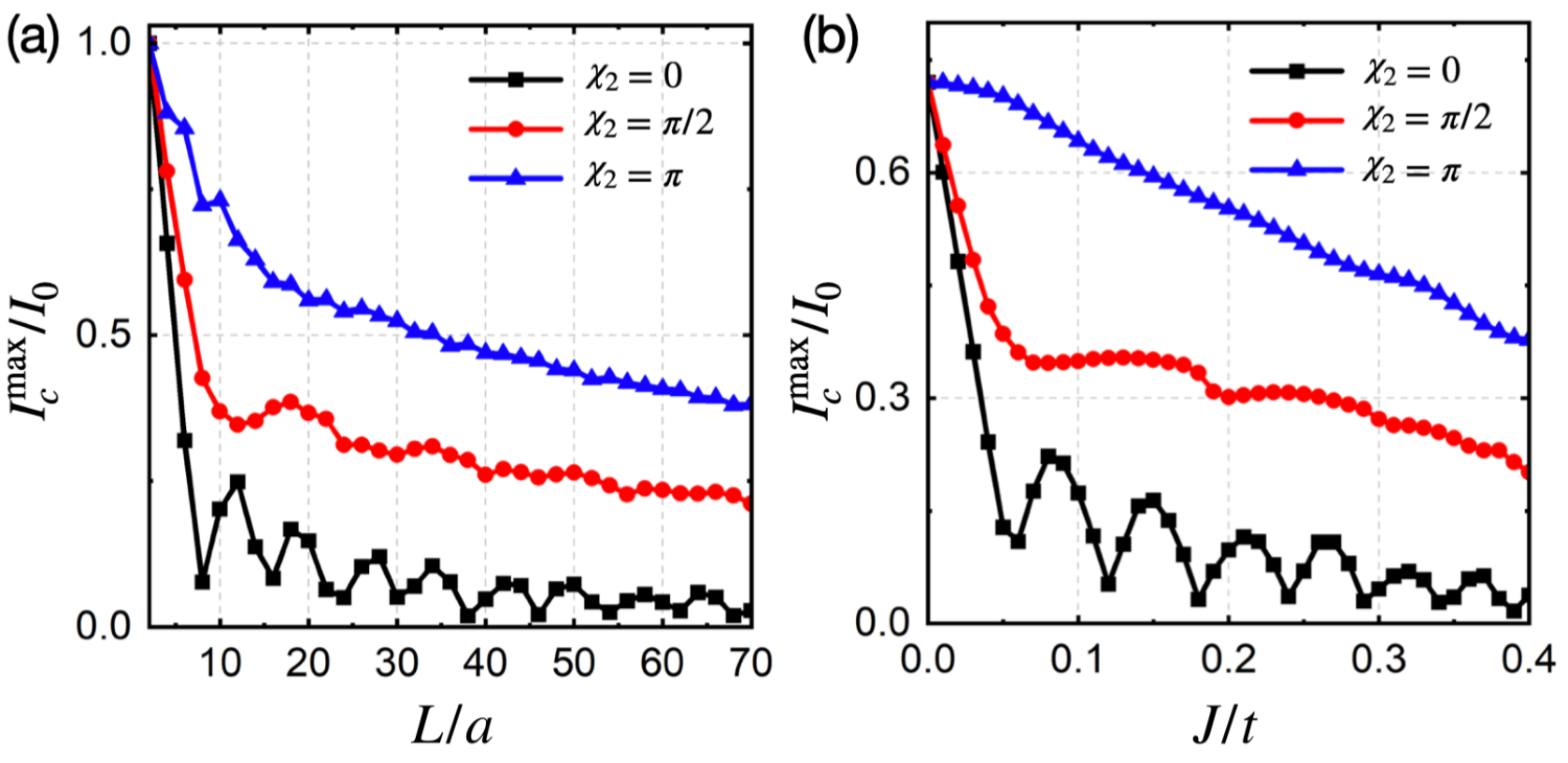}
\caption{Parameter dependence of the critical charge current $I_c^{\text{max}}$. (a) $I_c^{\text{max}}$ as a function of junction length $L$ (from $2a$ to $70a$) for $J=0.25t$. (b) $I_c^{\text{max}}$ as a function of $J$ for $L=32a$. Different colors denote various $\chi_2$ with $\chi_1=0$ fixed. }
\label{fig3}
\end{figure}

Next, we investigate in Fig.~\ref{fig3}(a) the critical current $I_{c}^{\text{{max}}}=\text{max}_{\varphi}[I_c(\varphi)]$ as a function of the length $L$. We fix $\chi_1 = 0$ and maintain equal thicknesses for both AM layers. The results reveal distinct transport behaviors depending on the relative orientation of the N\'{e}el vectors. For the parallel configuration, $\chi_2 = 0$, the critical current exhibits pronounced oscillatory behavior as $L$ increases, and its magnitude quickly decays toward zero. This behavior originates from the center-of-mass momentum of Cooper pairs~\cite{zhang2024,Bo2024} and
confirms the $0$-$\pi$ transition identified in the current-phase relation, where the destructive interference of positive and negative current contributions across different $k_y$ channels leads to a significant suppression of the total supercurrent. We note that the $I_c$ oscillations are robust to small variations of $\chi_1$ and $\chi_2$ around $\chi_1=\chi_2=0$.

For the antiparallel configuration, $\chi_2 = \pi$, the critical current displays a monotonic and relatively slow decay as $L$ increases, notably without the oscillatory behavior. 
In the antiparallel alignment the compensation of superconducting pair amplitude oscillations in the AM1-AM2 region restores conventional $0$-type behavior across all $k_y$ channels, thereby enhancing the Josephson current by preventing inter-channel cancellation. For the intermediate case $\chi_2 = \pi/2$, the critical current also decays as $L$ increases, but it maintains a higher magnitude than for the parallel configuration while exhibiting much weaker oscillations. In this case, the N\'{e}el vector misalignment mitigates the inter-channel cancellation observed in the parallel configuration. Therefore, we conclude that the misorientation of AMs can give rise to a longer-ranged Josephson current compared to that of identical AMs.

In Fig.~\ref{fig3}(b), we show the dependence of the critical current $%
I_{c}^{\text{max}}$ on the altermagnetic strength $J$, where the current
exhibits a general decaying behavior. To understand this suppression, we
introduce a continuum model derived from the tight-binding lattice model,
where the lattice altermagnetic strength $J$ corresponds directly to the
continuum parameter $\mathcal{J}$. This continuum system is described by the
BdG Hamiltonian
\begin{equation}
\mathcal{\check{H}}_{\bm{k}}=\left( 
\begin{array}{cc}
\hat{h}_{\bm{k}} & \hat{\Delta} \\ 
\hat{\Delta}^{\dag } & -\hat{h}_{-\bm{k}}^{\ast }%
\end{array}%
\right) +\mathcal{\check{M}}_{\bm{k}}^{\alpha },  \label{hami}
\end{equation}%
with $\hat{h}_{\bm{k}}=\hbar ^{2}\bm{k}^{2}/(2m)-\mu $, wave vector $%
\bm{k}=(k_{x},k_{y})$, and $\mu $ being the uniform chemical potential. $%
\mathcal{\check{M}}_{\bm{k}}^{\alpha =1\left( 2\right) }$ denotes the
altermagnetic exchange potential of AM1 (AM2) and takes the form $\mathcal{%
\check{M}}_{\bm{k}}^{\alpha }=\mathcal{D}_{\bm{k}}^{\alpha }\mathcal{\check{S%
}}\left( \chi _{\alpha }\right) $, with $\mathcal{D}_{\bm{k}}^{\alpha }=%
\mathcal{J}_{{1}}^{\alpha }k_{x}k_{y}+\mathcal{J}_{{2}}^{\alpha }\left(
k_{x}^{2}-k_{y}^{2}\right) /2$, $\mathcal{J}_{1}^{\alpha }=2\mathcal{J}\sin
2\chi _{\alpha }$ and $\mathcal{J}_{2}^{\alpha }=2\mathcal{J}\cos 2\chi
_{\alpha }$. Based on this continuum model for a $d_{x^{2}-y^{2}}$%
-altermagnet, the suppression of the current can be understood by noting
that the wavevectors of electrons forming Cooper pairs change as a function
of $\mathcal{J}$, namely, 
\begin{equation}
k_{e\pm }=\pm \sqrt{\frac{2m\mu \pm 2m\mathcal{J}k_{y}^{2}-\hbar
^{2}k_{y}^{2}}{\hbar ^{2}\pm 2m\mathcal{J}}}.  \label{eq:CP-wavevector}
\end{equation}
As $\mathcal{J}$ increases, a wider range of transverse $k_y$ channels become evanescent, i.e., $\left\vert k_{y}\right\vert >\sqrt{2m\mu /\left( 2m\mathcal{J}+\hbar ^{2}\right)}$ is reached in Eq.~(\ref{eq:CP-wavevector}). The superconducting pair amplitude for these channels decays significantly, impairing the supercurrent transport capability of the junction. Consequently, the total critical current is progressively suppressed as the strength of the altermagnetic order increases. Nevertheless, the remaining propagating channels continue to dictate the transport characteristics, manifesting as $0$-$\pi$ oscillations for $\chi_2 = 0$ and sustained enhancements for the $\chi_2 = \pi/2$ and $\chi_2 = \pi$ cases. 

\section{Josephson spin current}
\label{sec4}

Beyond charge transport, noncollinear altermagnetic heterostructures facilitate the emergence of exotic spin-polarized supercurrents. This phenomenon is rooted in the evolution of the Cooper pair spin wavefunction as it traverses the AM layers.
When a spin-singlet pair from the superconducting lead SC1 enters the altermagnetic layer AM1, the spin component of the wavefunction evolves from  $\frac{1}{\sqrt{2}}(\left|\uparrow\downarrow\right\rangle - \left|\downarrow\uparrow\right\rangle)$ to $\frac{1}{\sqrt{2}}(e^{iQ_{1,k_y} x}\left|\uparrow\downarrow\right\rangle_{\chi_1} - e^{-iQ_{1,k_y} x}\left|\downarrow\uparrow\right\rangle_{\chi_1})$. Here, $Q_{1,k_y} = k_{e+}^{\text{AM1}} + k_{e-}^{\text{AM1}}$ represents the center-of-mass momentum of the pair amplitude formed by electrons with opposite spins aligned with the N\'{e}el vector $\hat{n}_1$ in AM1. By projecting this state, we decompose the wavefunction into a spin-singlet component $|S\rangle$ and a spin-triplet component $|T_0\rangle_{\chi_1}$ as $\cos(Q_{1,k_y} x) |S\rangle + i \sin(Q_{1,k_y} x) |T_0\rangle_{\chi_1}$ with
\begin{equation}
   |T_0\rangle_{\chi_1} = \frac{1}{\sqrt{2}}(\left|\uparrow\downarrow\right\rangle_{\chi_1} +\left|\downarrow\uparrow\right\rangle_{\chi_1}). 
\end{equation}
The $|T_0\rangle_{\chi_1}$ state acts as the zero-energy eigenstate of the spin operator aligned with the local N\'{e}el vector $\hat{n}_1$. Consequently, the $\bf{d}$-vector characterizing this triplet state is intrinsically locked to the N\'{e}el vector, expressed as $\bf{d}_\alpha = (\cos \chi_\alpha, \sin \chi_\alpha, 0)$ for $\alpha \in \{1, 2\}$~\cite{Sigrist_Ueda_RMP, Keidel2014Aug}. As Cooper pairs transit the interface from AM1 to AM2, the shift from $\bf{d}_1$ to $\bf{d}_2$ requires a rotation of the spin-quantization axis. Such a relative misalignment $\Delta \chi = \chi_2 - \chi_1$ acts as a driving torque, generating a dissipationless and out-of-plane spin-polarized supercurrent.

Within the low-energy tunneling approximation, the momentum-resolved spin current $I_{s}(k_{y})$ can be  determined by the interference of the induced triplet anomalous Green's functions $\hat{f}_{\mathrm{t},\alpha}$.
The triplet $\bf{d}$-vector is parallel to the local effective exchange field ($\mathbf{d}_{\alpha} \parallel \hat{n}_{\alpha}$) so the triplet anomalous Green's functions can be expressed as $\hat{f}_{\mathrm{t},\alpha}= A_{\alpha,k_y} \hat{n}_\alpha \cdot \hat{\mathbf{s}}\, (i\hat{s}_y)$, with $\hat{\mathbf{s}}$ the vector of Pauli spin matrices and $A_{\alpha,k_y}$ a form factor. Consequently, evaluating the spin trace yields $I_{s}(k_{y}) \propto \mathrm{Re}(A_{1,k_y} A_{2,k_y}^{*}) (\hat{n}_{1} \times \hat{n}_{2})_{z}$, where the form factor $A_{\alpha}$ sensitively depends on the orientation of N\'{e}el vectors, momentum $k_y$ and superconducting phase.
This explicit dependence on $(\hat{n}_{1} \times \hat{n}_{2})_{z}$ demonstrates that the spin supercurrent vanishes for collinear N\'{e}el vectors, whereas a non-collinear orientation directly drives a net spin supercurrent.
This mechanism aligns with the established framework for spin supercurrent generation in Josephson junctions involving spin-triplet superconductors~\cite{asano2006triplet,Brydonprl2009,Manske2008,Brydonnjp2009}. While intrinsic spin-triplet superconductors can inherently support spin supercurrents, candidate triplet superconductors are exceptionally rare and their existence remains an open question~\cite{pustogow2019constraints}, thereby highlighting the crucial role of double-AM structures in providing a viable and tunable platform to realize dissipationless spin supercurrents using conventional $s$-wave superconductors.

\begin{figure*}[tb]
\centering
\includegraphics[width=15cm]{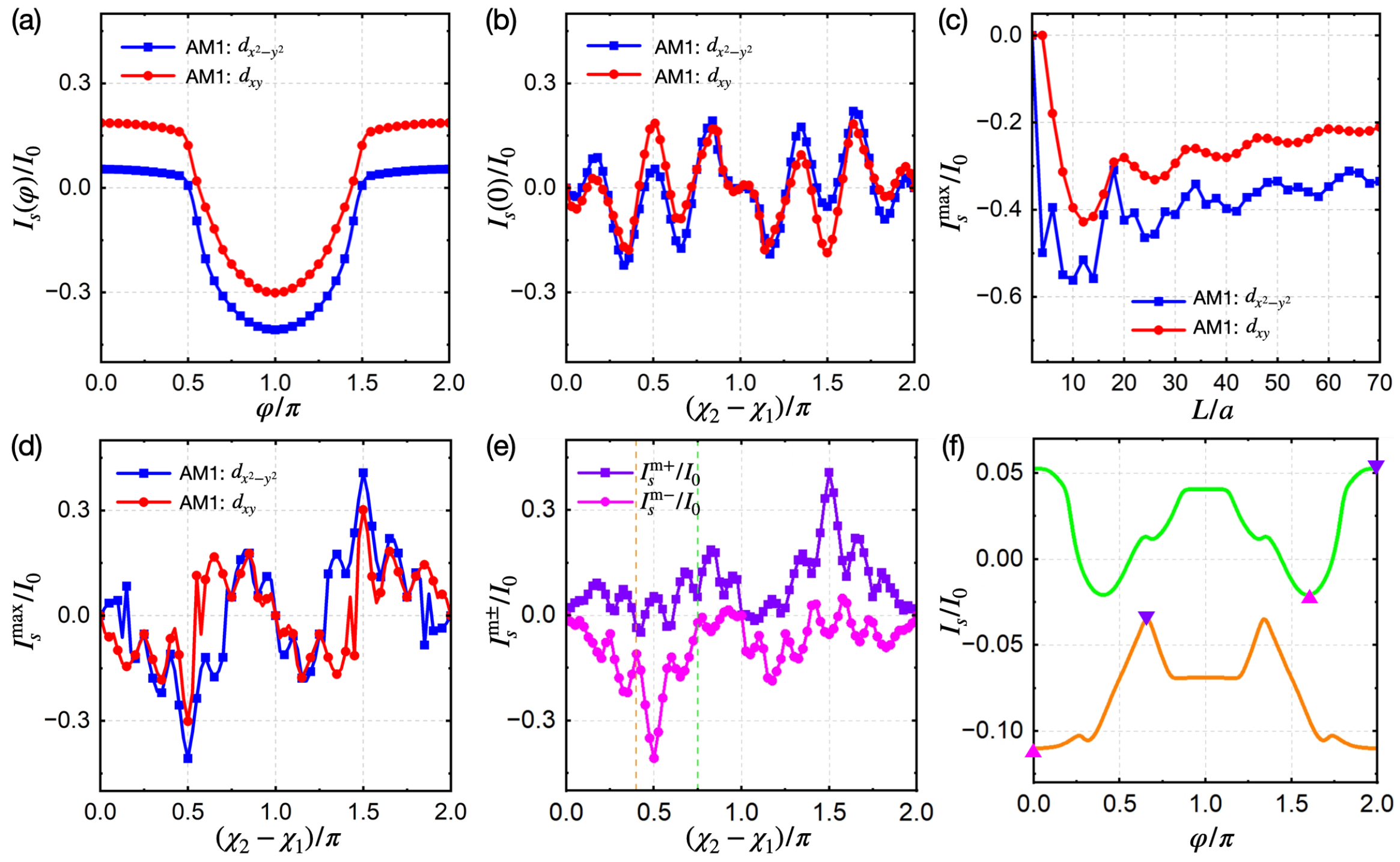}
\caption{(a) Spin supercurrent $I_s$ as a function of the superconducting phase difference $\varphi$ with a fixed crystal misorientation $\chi_2 - \chi_1 = \pi/2$. (b) Spin supercurrent $I_s$ at zero phase difference ($\varphi = 0$) as a function of the relative orientation $\chi_2 - \chi_1$. (c) Maximum spin supercurrent $I_s^{\text{max}}$ versus the junction length $L$ for $\chi_2 - \chi_1 = \pi/2$. (d) Maximum spin supercurrent $I_s^{\text{max}}$ as a function of the misorientation $\chi_2 - \chi_1$.
(e) Maximum value of the spin supercurrent $I_s^{\text{m}+}$ (violet) and minimum one $I_s^{\text{m}-}$ (magenta) as a function of $\chi_2 - \chi_1$ for the case corresponding to the blue curve in (d). (f) Spin supercurrent $I_s$ versus $\varphi$ for the parameter sets marked by the orange and green dashed lines in (e)}.
In all panels, we set $J=0.25t$ and the blue curves correspond to the $d_{x^2-y^2}$ case with $\chi_1 = 0$ and the red curves to the $d_{xy}$ case with $\chi_1 = \pi/4$. Except for panel(c), the junction length is fixed at $L = 22a$.
\label{fig4}
\end{figure*}

In Fig.~\ref{fig4}, we examine the generation of spin-polarized supercurrents, Eq.~(\ref{current}), in junctions featuring non-collinear N\'{e}el vectors. Figure \ref{fig4}(a) displays the dependence of the spin supercurrent $I_s$ on the superconducting phase difference $\varphi$ for the crystal misorientation $\chi_2 - \chi_1 = \pi/2$, comparing $\chi_1 = 0$ ($d_{x^2-y^2}$, {\color{blue}{blue curve}}) and $\chi_1 = \pi/4$ ($d_{xy}$, red curve). These results demonstrate that a finite spin current emerges in non-collinear configurations which is an even-function of the phase difference $\varphi$, i.e., $I_s(\varphi)=I_s(-\varphi)$. 
%{\color{blue}From a symmetry perspective, the combined operation $\mathcal{O} = \mathcal{T} R_z(\pi)$ leaves the spin current operator invariant, constraining the spin supercurrent to be an even function of $\varphi$~\cite{Sun2005}.} 
In analogy to a SC-FM-FM-SC junction with noncollinear magnetization~\cite{Shomali_2011}, we also observe a non-zero spin supercurrent at zero phase bias $\varphi = 0$. Figure \ref{fig4}(b) illustrates the spin supercurrent $I_s$ at zero phase difference as a function of the relative orientation $\chi_2 - \chi_1$. Both the magnitude and sign of $I_s(0)$ are highly sensitive to crystal misorientation, suggesting that altermagnetic junctions without a phase bias can serve as a source of spin supercurrent when the N\'{e}el vectors are non-collinear. 
In Fig.~\ref{fig4}(c), we show the maximum spin supercurrent $I_s^{\text{max}}$ as a function of the junction length $L$ for $\chi_1 = 0$ and $\chi_1 = \pi/4$ with a fixed relative angle $\chi_2 - \chi_1 = \pi/2$. 
Here, $I_{s}^{\text{{max}}}=I_s(\varphi_{p})$, with $\varphi_p = \arg\max_{\varphi \in [0,2\pi)} |I_s(\varphi)|$; that is, the spin current with maximum magnitude over the phase range $\varphi \in [0, 2\pi)$.
The results indicate that the spin supercurrent can sustain over a long range of junction length $L$ without significant decay.

Figure~\ref{fig4}(d) presents $I_s^{\text{max}}$ as a function of the crystal misorientation $\chi_2 - \chi_1$. 
Notably, and unlike the non-collinear ferromagnetic case~\cite{Shomali_2011}, the spin supercurrent depends on the specific values of both $\chi_1$ and $\chi_2$ rather than solely on their relative difference. 
To further elucidate the full angular dependence and the extremal bounds of the spin supercurrent, Fig.~\ref{fig4}(e) shows both the maximum [$I_s^{\text{m}+}=I_s(\varphi_{+})$, with $\varphi_+ = \arg\max_{\varphi \in [0,2\pi)} I_s(\varphi)$] and the minimum [$I_s^{\text{m}-}=I_s(\varphi_{-})$, with $\varphi_{-} = \arg\min_{\varphi \in [0,2\pi)} I_s(\varphi)$] values of the spin current against the relative orientation $\chi_2 - \chi_1$ for the $d_{x^2-y^2}$-wave symmetry, corresponding to the blue curve in Fig.~\ref{fig4}(d). The results demonstrate that the spin supercurrent vanishes at specific misorientation angles; namely, when $\chi_2 - \chi_1 = 0$, $\pi$, or $2\pi$. These points correspond to configurations where the N\'{e}el vectors achieve collinear alignment. Besides the collinear cases, a substantial spin supercurrent emerges, varying continuously between the established $I_s^{\text{m}+}$ and $I_s^{\text{m}-}$ bounds. 
Figure~\ref{fig4}(f) further details the phase dependence of the spin current, $I_s(\varphi)$, for two representative misorientation angles marked by the green and orange dashed lines in Fig.~\ref{fig4}(e). For the green curve, $I_s(\varphi)$ oscillates across zero, yielding opposite signs for $I_s^{\text{m}+}$ and $I_s^{\text{m}-}$. By contrast, for the orange curve both $I_s^{\text{m}+}$ and $I_s^{\text{m}-}$ share the same negative sign across the entire phase interval $\varphi \in [0, 2\pi)$. This single-signed behavior indicates that the spin current remains strictly unidirectional regardless of the phase bias, thereby enabling perfect nonreciprocal spin-polarized transport. Collectively, these findings demonstrate that inhomogeneous altermagnetic interlayers provide a highly tunable, field-free platform for generating and manipulating spin-polarized supercurrents.

\section{Conclusion}
\label{sec5}

In conclusion, we have investigated the Josephson transport in a hybrid junction with a composite altermagnetic interlayer. We demonstrated that the misorientation between the two altermagnetic N\'{e}el vectors significantly modulates both charge and spin supercurrents. At a $\pi$ misorientation, the superconducting pair amplitude oscillations, driven by center-of-mass momenta, undergo mutual cancellation in layers of equal thickness. This compensation suppresses pair-breaking, restores conventional transport, enhances the Josephson current, and eliminates $0$-$\pi$ transitions. Furthermore, the altermagnetic layers facilitate the conversion of spin-singlet pair amplitudes into spin-triplet ones. Ultimately, the twist of the $\bf{d}$-vectors generates a tunable, dissipationless spin-polarized current.

It is worth mentioning that our framework can also be applied to investigate systems with out-of-plane N\'{e}el vectors. In this case, since the crystal misorientation corresponds to a rotation within the $x-y$ plane, the spin polarizations remain strictly collinear regardless of the rotation angle. As the previous discussion shows, no spin-polarized supercurrent can then be generated. Furthermore, for out-of-plane N\'{e}el vectors, a relative angle of $\pi/2$ renders the two sub-lattice spin configurations anti-parallel which plays an equivalent role to the $\pi$-misorientation in the in-plane N\'{e}el vector scenario. Thus, instead of $\pi$-misorientation, the compensation effect for out-of-plane configurations is optimized at a $\pi/2$ misorientation where the spatial oscillation of the superconducting pair amplitude is perfectly canceled for equal layer thicknesses, ultimately leading to the most significant enhancement of the Josephson charge current.
Additionally, although our present framework focuses on the interplay between altermagnetism and superconductivity in the absence of spin-orbit coupling, a systematic investigation incorporating explicit spin-orbit interactions is an important direction for future research.

In terms of material candidates to experimentally test our predictions, it
might be possible to exploit recently reported $d$-wave AMs~\cite{JiangB2025}, e.g., metallic $\mathrm{Mn_5Si_3}$~\cite{MnSi}, $\mathrm{KV_2Se_2O}$~\cite{KVSeO}, and $\mathrm{Rb}_{1-\delta}\mathrm{V_2Te_2O}$~\cite{RbVTeO}. 
Moreover, recent research demonstrating the manipulation of altermagnetic order via electrical control~\cite{wang2025}, strain or composition engineering during film growth~\cite{fischer2026}, and field-cooling protocols~\cite{Amin2024} offers viable pathways to test our proposal.
Collectively, our results establish a field-free platform for generating spin-polarized Josephson currents, providing a robust framework for the development of functional superconducting spintronic devices.

\section{Acknowledgments}
Y.\ F.\ acknowledges financial support from the Sumitomo Foundation and JSPS with Grants-in-Aid for Scientific Research (KAKENHI Grant No.\ 26K17096).
P.\ B.\ acknowledges support by the Spanish CM ``Talento Program'' project No.~2019-T1/IND-14088 and No.~2023-5A/IND-28927, the Agencia Estatal de Investigaci\'on project No.~PID2020-117992GA-I00, No.~PID2024-157821NB-I00 and No.~CNS2022-135950 and through the Grant ``Excelencia María de Maeztu'' CEX2023-001316-M funded by MICIU/AEI/10.13039/501100011033. 
J.\ C.\ acknowledges financial support from the Swedish Research Council  (Vetenskapsr\aa det Grant No.~2021-04121), the Carl Trygger’s Foundation (Grant No. 22: 2093) and  the Olle Engkvist Foundation (Grant No.~243-1026). 
Y.\ T. \ acknowledges financial support from JSPS with Grants-in-Aid for Scientific Research (KAKENHI Grants Nos. 23K17668, 24K00583, 24K00556, 24K00578, 25H00609 and 25H00613). 
B.\ L.\  acknowledges financial support from the National Natural Science Foundation of China (project 12474049) and Beijing National Laboratory for Condensed Matter Physics (2025BNLCMPKF011).

\bibliography{altermagnet}

\end{document}